# Entropy production in temperature modulated differential scanning calorimetry

J.-L. Garden, J. Richard

*Institut Néel, CNRS et Université Joseph Fourier, BP 166, 38042 Grenoble Cedex 9, France.*

and

Y. Saruyama

*Division of Macromolecular Science and Engineering, Kyoto Institute of Technology, Matsugasaki, Sakyo, Kyoto 606-8585, Japan.*

*The non-equilibrium process due to irreversible heat exchanges occurring during a temperature modulated differential scanning calorimetry experiment is investigated in detail. This enables us to define an experimental frequency dependent complex heat capacity from this calorimetric method. The physical meaning of this dynamic heat capacity is discussed. A relationship is clearly established between the imaginary part of this complex quantity and the net entropy created during the experimental time-scale.*

Keywords: TMDSC; Frequency dependent complex heat capacity; Non-equilibrium thermodynamics; Entropy production; Relaxation time constants.

\* Corresponding author. Fax: 33 (0) 4 76 87 50 60

*E-mail address:* jean-luc.garden@grenoble.cnrs.fr (J.-L. Garden)

# 1. Introduction

Gobrecht *et al.* had the original idea to add a temperature oscillation to the usual temperature program of a differential scanning calorimeter [1]. Later, Gill and co-workers refund this principle now called the temperature modulated differential scanning calorimetry (TMDSC) [2]. Now, this differential calorimetric method is used as often as the classical differential scanning calorimetry (DSC). By means of a deconvolution, Gill has proposed to separate the measured signal into a reversing and non-reversing component. After, as in the well-known $3\omega$-method of Birge and Nagel [3], Schawe has proposed to separate the two components of the TMDSC signal in a real and imaginary part for the measured dynamic heat capacity [4]. This generalized calorimetric susceptibility is the consequence of the slow decay close to equilibrium of internal degrees of freedom which cannot instantaneously follow the oscillating temperature rate imposed by the experimentalist (see reference [5] and references therein). As shown recently, when an experimental time constant intervenes in a calorimetric modulated temperature experiment such as the ac-calorimetry method, then the measured heat capacity becomes a dynamic quantity which is also represented by a complex number [6]. In the present paper, we call these complex quantities "experimental frequency dependent complex heat capacities" in order to make a clear distinction between these dynamic quantities and the usual frequency dependent complex heat capacity that we mostly call generalized calorimetric susceptibility. The imaginary parts of these experimental frequency dependent complex heat capacities are always connected to the net entropy produced during the experimental time scale like in the case of the generalized calorimetric susceptibility. This creation of entropy is always due to a particular physical irreversible process for which the relaxation time constant plays a major role.

In this paper, it is proposed to tackle only the non-equilibrium process due to irreversible heat exchanges between the different parts of a usual TMDSC calorimetric head. It is shown that an experimental frequency dependent complex heat capacity can be inferred from these heat exchanges. The diffusion of heat within the different parts of the calorimeter, due to slow thermal diffusivity or bad thermal contact as well as relaxation of the heat carriers within each cell (see for details reference [6]) are not considered here. In the part two, the physical meaning of this experimental frequency dependent complex heat capacity is discussed. In the part three, the entropy due to the internal heat exchanges produced during one period of a temperature cycle is calculated.

## 2. Experimental frequency dependent complex heat capacity in TMDSC.

*2.1 Notations and assumptions*

Let us consider a schematic TMDSC calorimetric head as depicted in the figure 1. Two calorimetric cells of same addenda heat capacity, one for the sample and the other for the reference are thermally connected to a cell-holder through two identical heat exchange coefficients $K$. The thermal link which connects the two cells each others is supposed negligible and neglected as compared to $K$. Hence, the two cells are thermally isolated each others. The cell-holder is connected to the real thermal bath of constant temperature $T_0'$ via another thermal link $K_0$. This latter thermal link does not play any role in ours calculations because we consider only the thermodynamic system constituted by the two calorimetric cells and the cell-holder. The boundary of the system is represented by an ellipsis in the figure 1. The cell-holder is a thermal bath for the temperature of each cell. Its heat capacity is thus much higher than that of each cell even when they contain the sample and the reference. The differential temperature is represented in the figure 1 by a thermocouple with two junctions of different metals, but we have to bear in mind that it could be any thermometric device settled in differential way such as a thermopile constituted by many thermocouples connected in series or by two different thermometers mounted in opposition in a Wheatstone bridge, depending on the TMDSC calorimeter. The thermal power necessary to produce the ramp and the harmonic oscillation is supplied directly to the cell-holder. This is represented by an arrow in the figure 1. If the thermal power is supplied directly to the cells, then we are in presence of differential ac-calorimetry experiment. This is the principal difference between the two methods. The thermal power is generally supplied by a heater (Joule effect) and the temperature of the cell-holder (thermal bath) is recorded by a thermometer. These two sensitive elements are not represented in the figure 1. The temperature of the bath can be decomposed in a dc and an ac component:

$$T_b = T_0 + T_b^{ac} \tag{1}$$

with $T_b^{ac} = \widetilde{T}_b \exp(i\omega t)$. The phases of the oscillating temperatures of the two cells are referenced to the one of the bath (taken by convention equal to zero). In a real experiment, the dc or mean temperature $T_0$ is included in a servo-system in order to follow a perfect ramp. In this paper, we consider only the stationary condition for which $T_0$ is maintained constant in the course of time. In fact, if we take into account this ramp the traditional DSC case is refund and it brings us no new information for our demonstration. In other words, we have supposed that a perfect ramp regime is attained and that all our differential equations take into account only the oscillating parts in reference to this stationary ramp state. The temperature of the sample is:

$$T_s = T_s^{dc} + T_s^{ac} \qquad (2)$$

with $T_s^{ac} = \widetilde{T}_s \exp i(\omega t - \varphi_s)$. The temperature of the reference is:

$$T_r = T_r^{dc} + T_r^{ac} \qquad (3)$$

with $T_r^{ac} = \widetilde{T}_r \exp i(\omega t - \varphi_r)$. As already mentioned in the introduction, the temperature gradients inside each component of the system are neglected. Also, the temperature gradients due to bad thermal contacts between each element such as thermal interfaces between the heaters or the thermometers and the different elements of the system are neglected. Only thermal gradients due to the heat exchange coefficients $K$ are considered. In other words, each cell and the cell-holder are homogeneous in temperature. Finally, an important assumption is to consider that the dc temperature difference or the ac amplitudes of temperature modulations are negligible in relation to the absolute temperature of each element:

$$\Delta T_i^{dc}, \widetilde{T}_i^{ac} \ll T_0, T_i^{dc} \text{ with } i = s, r \qquad (4)$$

This last hypothesis allows us to linearize the future differential equations. On a non-equilibrium thermodynamics point of view, this is the hypothesis of the linear regime.

*2.2 TMDSC calorific equations*

A thermal power is supplied to the cell-holder in such a way that its temperature follows a linear ramp plus an oscillation at a well determined frequency. The cell-holder constitutes a thermal bath for the temperature of the cells because its heat capacity is much greater than those of the cells. In the following we will make no distinction between the cell-holder and the thermal bath. The temperature of the two cells obeys to the two following first order differential equations:

$$\begin{cases} C_s \dot{T}_s = -K(T_s - T_b) \\ C_r \dot{T}_r = -K(T_r - T_b) \end{cases} \quad (5)$$

where $C_s$ and $C_r$ are the total heat capacities (addenda plus sample or reference) of the sample-cell and reference-cell respectively. From the initial notations (1), (2) and (3), these two equations can be both separated into dc and ac equations:

$$\begin{cases} C_s \dot{T}_s^{dc} = -K(T_s^{dc} - T_0) \\ C_r \dot{T}_r^{dc} = -K(T_r^{dc} - T_0) \end{cases} \quad (6)$$

$$\begin{cases} C_s \dot{T}_s^{ac} = -K(T_s^{ac} - T_b^{ac}) \\ C_r \dot{T}_r^{ac} = -K(T_r^{ac} - T_b^{ac}) \end{cases} \quad (7)$$

The system (6) is just the classical differential equations of the DSC and as already mentioned it is not considered in this paper. The authors are aware that it might be interesting to treat the classical DSC case under the point of view of non-equilibrium thermodynamics when irreversible effects take place. For information on such a treatment see the reference [5]. The resolution of (7) in the stationary regime (thus compared to the ramp state) gives:

$$\begin{cases} T_s^{ac} = \dfrac{T_b^{ac}}{1+i\omega\tau_s} \\ T_r^{ac} = \dfrac{T_b^{ac}}{1+i\omega\tau_r} \end{cases} \quad (8)$$

with $\tau_s = C_s/K$ and $\tau_r = C_r/K$ the two relaxation time constants of the temperatures of both cells. This simply indicates that at zero frequency, the temperatures of the two cells and the temperature of the bath oscillate in phase, and at infinite frequency the temperatures of the cells never oscillate anymore.

*2.3 Experimental frequency dependent complex heat capacity*

By analogy with the ac-calorimetry method where the complex heat capacity is defined as the ratio of the oscillating thermal power supplied to the sample over the oscillating temperature rate:

$$C^* = \frac{P_{ac}}{\dot{T}_{ac}} = \frac{P_{ac}}{i\omega T_{ac}} \quad (9)$$

it is also possible to define a complex heat capacity in TMDSC. We have however to bear in mind that in TMDSC the oscillating power is supplied onto the thermal bath. Thus, it is the oscillating temperature of the thermal bath that we have take into account in the ratio. Consequently, the complex heat capacity is defined as:

$$C_s^* = \frac{P_s^{ac}}{\dot{T}_b^{ac}} = \frac{P_s^{ac}}{i\omega T_b^{ac}} \tag{10}$$

for the sample-cell only, and where $P_s^{ac}$ is the oscillating thermal power at the level of the thermal bath which can be expressed as follow:

$$P_s^{ac} = C_b \dot{T}_b^{ac} + C_s \dot{T}_s^{ac} \tag{11}$$

For the reference cell we have identically:

$$C_r^* = \frac{P_r^{ac}}{\dot{T}_b^{ac}} = \frac{P_r^{ac}}{i\omega T_b^{ac}} \tag{12}$$

with:

$$P_r^{ac} = C_b \dot{T}_b^{ac} + C_r \dot{T}_r^{ac} \tag{13}$$

In fact, our reasoning is based on two different experiences, one where only the sample-cell is present and the other where only the reference-cell is present as depicted in the figure 2. The exact result of the TMDSC is the difference between these two "mono-experiments". From (10),(11) and (8) we obtain:

$$C_s^* = C_b + \frac{C_s}{1+i\omega\tau_s} \tag{14}$$

With the same type of formula for the reference:

$$C_r^* = C_b + \frac{C_r}{1+i\omega\tau_r} \tag{15}$$

We can discuss at this point the physical meaning of (14) or (15). When the thermal frequency of the input ac power is small as compared to $1/\tau_s$ ($\omega\tau_s \ll 1$) then the real heat capacity $C_b + C_s$ of the whole (heat bath plus sample) is measured. On the opposite, when the inequality $\omega\tau_s \gg 1$ holds then only the heat bath heat capacity is measurable. The equations (14) or (15) can be compared by analogy to the usual equation of the generalized calorimetric susceptibility (see Ref. [5] and [7]) where at low frequency all the internal degrees of freedom participate to the heat capacity measurement, and at high frequency only the fast degrees of freedom (phonon bath) are measured, the other slow modes being frozen-in. By making the difference between (14) and (15) the differential temperature measurement provides directly:

$$\Delta C^* = \frac{C_s}{1+i\omega\tau_s} - \frac{C_r}{1+i\omega\tau_r} \tag{16}$$

which is the expression of the experimental differential frequency dependent complex heat capacity in TMDSC where only internal oscillating irreversible exchanges of heat have been considered. Obviously, we could have considered other irreversible effects such as heat diffusion within the different media and particularly relaxation of some slow internal degrees of freedom. This latter effect modifies directly the heat capacity $C_s$ at the numerator of (16) following the usual definition of the generalized calorimetric susceptibility.

**3. Entropy production in TMDSC**

In the figure 1, the total variation of entropy involved in the entire differential system delimited by the ellipsis is:

$$dS_{tot} = \frac{\delta Q^s}{T_s} + \frac{\delta Q^r}{T_r} + \frac{\delta Q^b}{T_b} \qquad (17)$$

The $\delta Q^j$ ($i = s,r,b$) are the heat exchanges (to or from) the three different bodies of the differential calorimetric head. In this system, $dQ^s = dQ_i^s$ and $dQ^r = dQ_i^r$ meaning that heat exchanges at the level of the sample or the reference are only internal exchanges. We have also the evident two following equations: $dQ_i^s + dQ_i^{b/s} = 0$ and $dQ_i^r + dQ_i^{b/r} = 0$ which means that heat exchanges at the level of the sample and the reference compulsorily come from the heat bath (which is as mentioned in the forgoing one of the principal feature of the TMDSC). At the level of the bath, we have: $dQ^b = dQ_i^{b/s} + dQ_i^{b/r} + dQ_e^b$ where there is an external contribution coming from the surroundings. This external contribution is the quantity of heat supplied at the level of the bath via a heater in order to produce a linear ramp and an oscillating temperature at the level of the sample-cell and the reference-cell. With these different equations and notations the total entropy variation of the system is written:

$$dS_{tot} = \frac{dQ_i^s}{T_s} + \frac{dQ_i^r}{T_r} + \frac{dQ_i^{b/s}}{T_b} + \frac{dQ_i^{b/r}}{T_b} + \frac{dQ_e^b}{T_b} \qquad (18)$$

which can be separated into an internal and an external contributions. The internal contribution is explicitly written:

$$d_i S = dQ_i^s \left[ \frac{1}{T_s} - \frac{1}{T_b} \right] + dQ_i^r \left[ \frac{1}{T_r} - \frac{1}{T_b} \right] \qquad (19)$$

where we recognize the internal production of entropy written as a product of a thermodynamic force (difference of the inverse of the temperatures) with a thermodynamic flux (heat flux if we take the time derivative) as it is well-known in non-equilibrium thermodynamics. We might notice at this level the following remark: since in TMDSC the measurement is differential (the temperature difference between the sample-cell and the reference-cell is directly measured) then the differential heat capacity (equation (16)) is directly obtained. Thus, in order to establish a possible comparison between the imaginary part of the differential complex heat capacity and the entropy production in TMDSC, it is necessary to consider the differential entropy production. Consequently, the amount of heat exchanged between the sample-cell and the bath should be considered with an opposite sign as compared to the amount of heat exchanged between the reference-cell and the bath (or vice-versa). The differential internal entropy production in TMDSC is thus written:

$$d_i S = dQ_i^s \left[ \frac{1}{T_s} - \frac{1}{T_b} \right] - dQ_i^r \left[ \frac{1}{T_r} - \frac{1}{T_b} \right] \qquad (20)$$

Thanks to the assumption made in the section 1 where it has been considered that the temperature differences are not too high as compared to the absolute temperatures, this last expression can be linearized. For example, we have:

$$\frac{1}{T_s} - \frac{1}{T_b} \approx -\left[\frac{\Delta T_s^{dc} + T_s^{ac} - T_b^{ac}}{(T_0)^2}\right] \tag{21}$$

where we have written $\Delta T_s^{dc} = T_s^{dc} - T_0$. A similar expression is obtained for the reference. We can easily separate the dc and the ac contributions, but as already mentioned only the ac contributions are considered. Consequently, the ac contribution of the instantaneous differential entropy production is written as follows:

$$\sigma_i^{ac} = \frac{d_i S}{dt} = -\frac{dQ_i^s}{dt}\left[\frac{T_s^{ac} - T_b^{ac}}{T_0^2}\right] - \left\{-\frac{dQ_i^r}{dt}\left[\frac{T_r^{ac} - T_b^{ac}}{T_0^2}\right]\right\} \tag{22}$$

The expressions of the ac heat flux induced by the difference of the inverse of the temperatures in TMDSC are simply given by (7) which is rewritten here for the sake of clarity:

$$\begin{cases} \dfrac{dQ_i^s}{dt} = -K(T_s^{ac} - T_b^{ac}) \\ \dfrac{dQ_i^r}{dt} = -K(T_r^{ac} - T_b^{ac}) \end{cases} \tag{23}$$

They are simply the ac heat fluxes passing across the two identical heat exchange coefficient $K$. Accordingly, the instantaneous differential ac entropy production is:

$$\sigma_i^{ac} = \frac{K(T_s^{ac} - T_b^{ac})^2}{T_0^2} - \frac{K(T_r^{ac} - T_b^{ac})^2}{T_0^2} \tag{24}$$

It may be pointed out that this equation represents the difference of two entropies but taken at a second order with respect to the relative temperature difference. The entropy production is always a second order property in relation with the force or the flux and consequently in the linear regime close to equilibrium it represents always a small quantity. It can be also noticed that if there is no thermal event in the sample-cell, thus inevitably $T_s^{ac} = T_r^{ac}$ and the differential entropy production due to the oscillating parts of the temperature is equal to zero. At contrary, if there is a thermal event in the sample-cell during the process inducing a heat capacity change, the instantaneous differential entropy production becomes different from zero. In this case, if we take the integral of this last expression over one period of the temperature cycle, the following expression is finally obtained:

$$\overline{\sigma}_i^{ac} = \pi \frac{K}{\omega} \left( \frac{\tilde{T}_b}{T_0} \right)^2 \left[ \frac{1}{1+(\omega \tau_r)^2} - \frac{1}{1+(\omega \tau_s)^2} \right] \qquad (25)$$

where the two relaxation time constants appear. Now, knowing that the expression (16) obtained for the differential frequency dependent complex heat capacity can be also written as follows:

$$\Delta C^* = -\frac{K}{i\omega} \left[ \frac{1}{1+i\omega \tau_s} - \frac{1}{1+i\omega \tau_r} \right] = \Delta C' - i\Delta C'' \qquad (26)$$

with

$$\Delta C'' = \frac{K}{\omega} \left[ \frac{1}{1+(\omega \tau_r)^2} - \frac{1}{1+(\omega \tau_s)^2} \right] \qquad (27)$$

then, the well-known following relationship is obtained:

$$\overline{\sigma}_i^{ac} = \pi \left(\frac{\widetilde{T}_b}{T_0}\right)^2 \Delta C'' \tag{28}$$

Thus, when considering only the irreversible effect due to heat exchanges inside the different parts of the calorimetric head, a comparison is possible between the TMDSC and the ac-calorimetry [6]. The principal difference is that in the ac-calorimetry case, it is the amplitude of the temperature oscillation of the sample which appears directly in the relationship although in the TMDSC case it is the amplitude of the oscillation of the bath. This difference obviously comes from the fact that the perturbing thermal power is supplied to different levels in the two calorimetric methods.

## 4. Conclusion

In this paper, considering only the non-equilibrium effect due to irreversible internal heat exchanges inside a TMDSC calorimetric head, it is possible to obtain the following features:

-1. considering only the oscillating part of the differential temperature, an experimental frequency dependent complex heat capacity can be defined. This dynamic heat capacity has in fact the same expression than the usual one where the slow internal modes within a sample are taken into account. Obviously, in the two cases the physical meaning of these complex quantities is different. In the present case, the fact that the measured heat capacity is complex signifies that depending on the value of the thermal frequency, the heat capacity of the sample must be recorded or not. At sufficient low frequency the differential heat capacity between the sample and the reference can be recorded. This fact is well-known by all specialists of TMDSC. At high frequency, the differential signal gives a differential heat capacity equal to zero.

-2. the net entropy created during one period of the temperature cycle can be calculated in a differential mode (difference of the entropy produced by heat exchanges between the sample-cell and the cell-holder and the reference-cell and the cell-holder respectively).

-3. a clear relationship between this entropy variation and the imaginary part of the experimental frequency dependent differential complex heat capacity has been established. This relationship means that the imaginary part of $\Delta C^*$ is linked to an amount of heat which is not involved in the TMDSC differential heat capacity measurement. The higher the frequency, the greater the quantity of heat irreversibly lost for the measurement during one cycle. Let bear in mind that it is exactly the opposite case in ac-calorimetry for which the lower the frequency, the greater this quantity of heat lost via the heat leak (non-adiabaticity).

Knowing exactly the expression of this experimental frequency dependent complex heat capacity, it is now possible to take into account this unwanted complex effect in order to extract the real and imaginary parts of the dynamic heat capacity as required by Schawe when the kinetic of some internal degree of freedom plays a role [4]. Hence, under this point of view, it is possible to enlarge the thermal frequency range in TMDSC up to higher frequencies. However, we must remember that there exist an other irreversible effect which might perturb in a similar way the TMDSC measurement at high frequency. Indeed, bad thermal contact between the sample or reference and the cells or bad thermal contact between sensitive elements (thermometers and heaters) and each cell, or low intrinsic thermal diffusivity of the studied samples, can be limiting factors for the extraction of the physical property of the generalized calorimetric susceptibility.

Figure 1: A schematic view of a TMDSC calorimetric head is represented. Two cells of same heat capacities are thermally connected to a sample holder by two identical heat exchange coefficients. The cell-holder which is considered as a heat bath for the temperature of the two cells is connected to the real thermal bath of the calorimeter at constant temperature $T_0^{'}$. The thermal power necessary to produce the temperature ramp and the temperature oscillation of the cells is supplied directly at the level of the cell-holder. The differential temperature is schematically represented by a thermocouple. The thermodynamic system that is considered is only constituted by the two cells and the cell-holder (thermal bath).

Figure 2: In order to consider the differential measurement under a non-equilibrium thermodynamics point of view, the TMDSC experiment is seen as a difference between two single experiments, one with the sample-cell and the cell-holder and the other with the reference cell and the cell-holder.

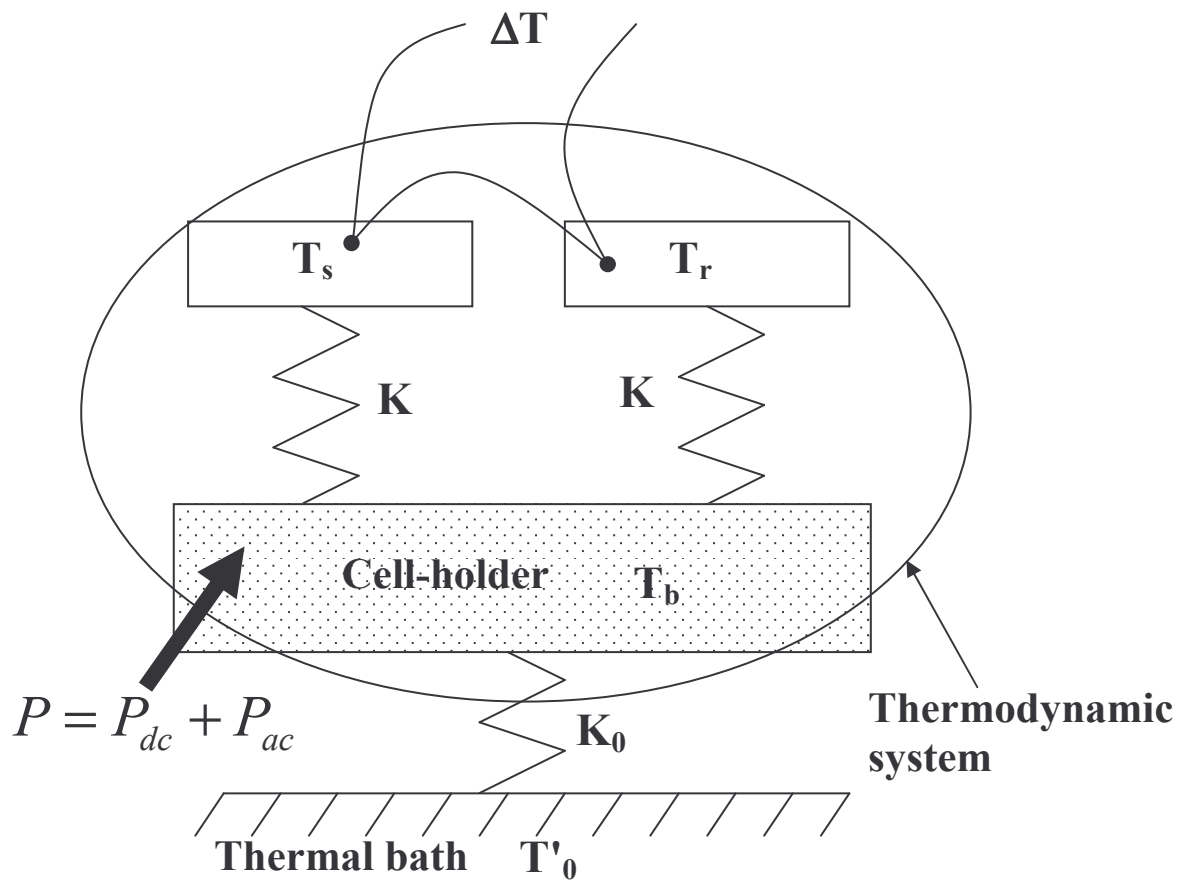

J.-L. Garden, Entropy production in temperature modulated differential scanning calorimetry, Figure 1
19

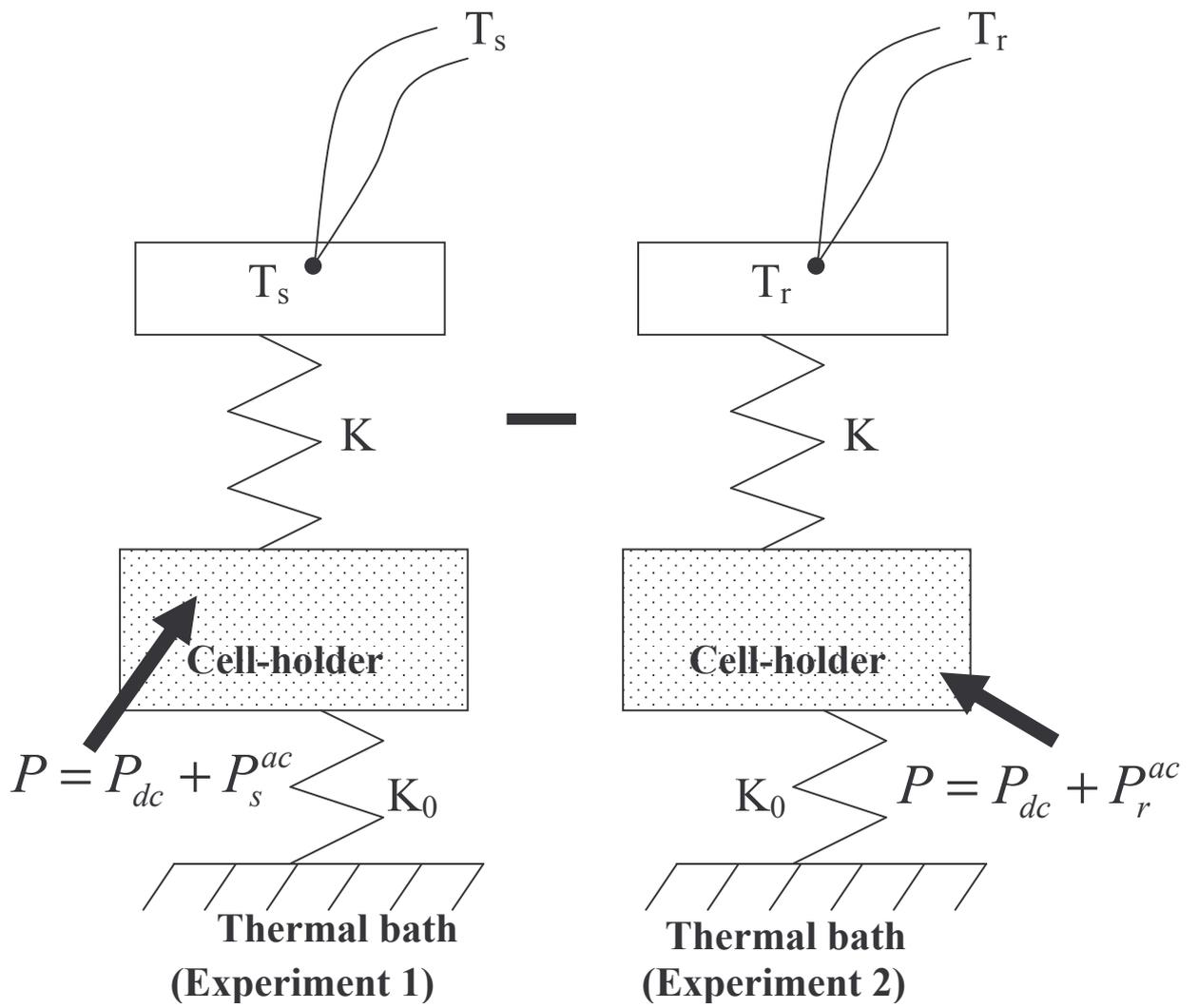

J.-L. Garden, Entropy production in temperature modulated differential scanning calorimetry, Figure 2